\documentclass[12pt]{article}

\usepackage{graphicx}
\usepackage{dcolumn}
\usepackage{bm}

\usepackage{graphicx}
\usepackage{url}

\newcommand\pubdate{\today}

\def\institute{Helsinki Institute of Physics\\
University of Helsinki, FINLAND}

\def\Title#1{\begin{center} {\Large #1 } \end{center}}
\def\Author#1{\begin{center}{ \sc #1} \end{center}}
\def\Address#1{\begin{center}{ \it #1} \end{center}}

\newcommand\pubblock{\rightline{\begin{tabular}{l}\\
         \pubdate  \end{tabular}}}
\newenvironment{Abstract}{\begin{quotation}  }{\end{quotation}}
\newenvironment{Presented}{\begin{quotation} \begin{center}
             PRESENTED AT\end{center}\bigskip
      \begin{center}\begin{large}}{\end{large}\end{center} \end{quotation}}

\def\beq{\begin{equation}}
\def\eeq#1{\label{#1}\end{equation}}
\def\eeqn{\end{equation}}

\def\beqa{\begin{eqnarray}}
\def\eeqa#1{\label{#1}\end{eqnarray}}
\def\eeqan{\end{eqnarray}}

\let\bar=\overbar

\def\O{{\cal O}}

\def\Dslash{\not{\hbox{\kern-4pt $D$}}}
\def\dslash{\not{\hbox{\kern-2pt $\del$}}}

\def\msb{{\bar{\ssstyle M \kern -1pt S}}}

\usepackage[utf8]{inputenc}
\usepackage[T1]{fontenc}
\usepackage{units}

\begin{document}
\begin{titlepage}
\pubblock

\vfill
\Title{Top mass shift resulting from the recalibration of flavor-dependent
jet energy corrections in the DØ lepton+jets top mass measurement}
\vfill
\Author{ \underline{Hannu Siikonen}, T. Mäkelä and M. Voutilainen }
\Address{\institute}
\vfill
\begin{Abstract}
We review the possible effects of the miscalibration of the flavor-dependent Jet Energy Corrections ($F_{\mathrm{Corr}}$) on the D\O\ lepton+jets top mass measurement.
The work is based on a previous study, where the D\O\ $F_{\mathrm{Corr}}$ calibration procedure was repeated based on the release of the internal D\O\ notes after a 5-year moratorium.
The cited study was motivated by the extraordinary precision claimed by D\O{} in their top mass measurement.
Using a Pythia6-based $F_{\mathrm{Corr}}$ recalibration, the $m_t$ result was shifted from 174.95 to 173.16 GeV.
Moreover, utilizing a Herwig7-based $F_{\mathrm{Corr}}$ calibration (not accounted for in the D\O{} studies), a shift down to 171.84 GeV was observed.
We find this both convincing and specific evidence for re-reviewing a part of the $F_{\mathrm{Corr}}$ calibration process.
However, D\O{} has been unwilling to open such studies or to provide other convincing counter-evidence.
\end{Abstract}
\vfill
\begin{Presented}
$13^\mathrm{th}$ International Workshop on Top Quark Physics\\
Durham, UK (videoconference), 14--18 September, 2020
\end{Presented}
\vfill
\end{titlepage}
\def\thefootnote{\fnsymbol{footnote}}
\setcounter{footnote}{0}

\section{Introduction}

The most notable measurements of the top quark mass ($m_t$) are those performed by the Tevatron collaborations CDF and D\O{} and those performed by the LHC collaborations ATLAS and CMS.
The D\O{} $m_t$ combination ($174.95 \pm 0.75$~GeV) \cite{ref:d0comb} is a distinct outlier in the $m_t$ world combination \cite{pdg}.
The measurements by CDF, ATLAS and CMS tend to produce values ranging from 172 to 173 GeV.
Moreover, the D\O{} combination is almost completely driven by the lepton+jets measurement ($174.98\pm 0.76$~GeV) \cite{ref:d02015}.
This gives motivation for studying the calibrations behind this measurement.

The Master's thesis written by Toni Mäkelä \cite{ref:tonimaster} made an in-depth study of the D\O{} flavor-dependent Jet Energy Corrections ($F_{\mathrm{Corr}}$).
It was based on the D\O{} internal notes that were released from their five-year moratorium early in 2018.
Some D\O{} authors were helpfully collaborating with Toni, and provided him with all the necessary material and communication.
As a notable detail, the D\O\ Run~IIb $F_{\mathrm{Corr}}$'s differ considerably from those of Run~IIa.
Toni's most significant finding was that the Run~IIb results should have indeed resembled more those of Run~IIa.

In the $m_t$ shifting analysis, performed by the main author of this text, we apply the $F_{\mathrm{Corr}}$ values produced by Toni into a lepton+jets top mass measurement.
The study utilizes a standalone top quark pair simulation and a detailed mathematical framework for propagating the $F_{\mathrm{Corr}}$ changes into the $m_t$ measurement \cite{siikonen:2020mass}.
In the TOP2020 poster session, the results obtained with this method were presented.

\begin{figure}[ht!] \label{fig:toni}
\centering
\includegraphics[width=\linewidth]{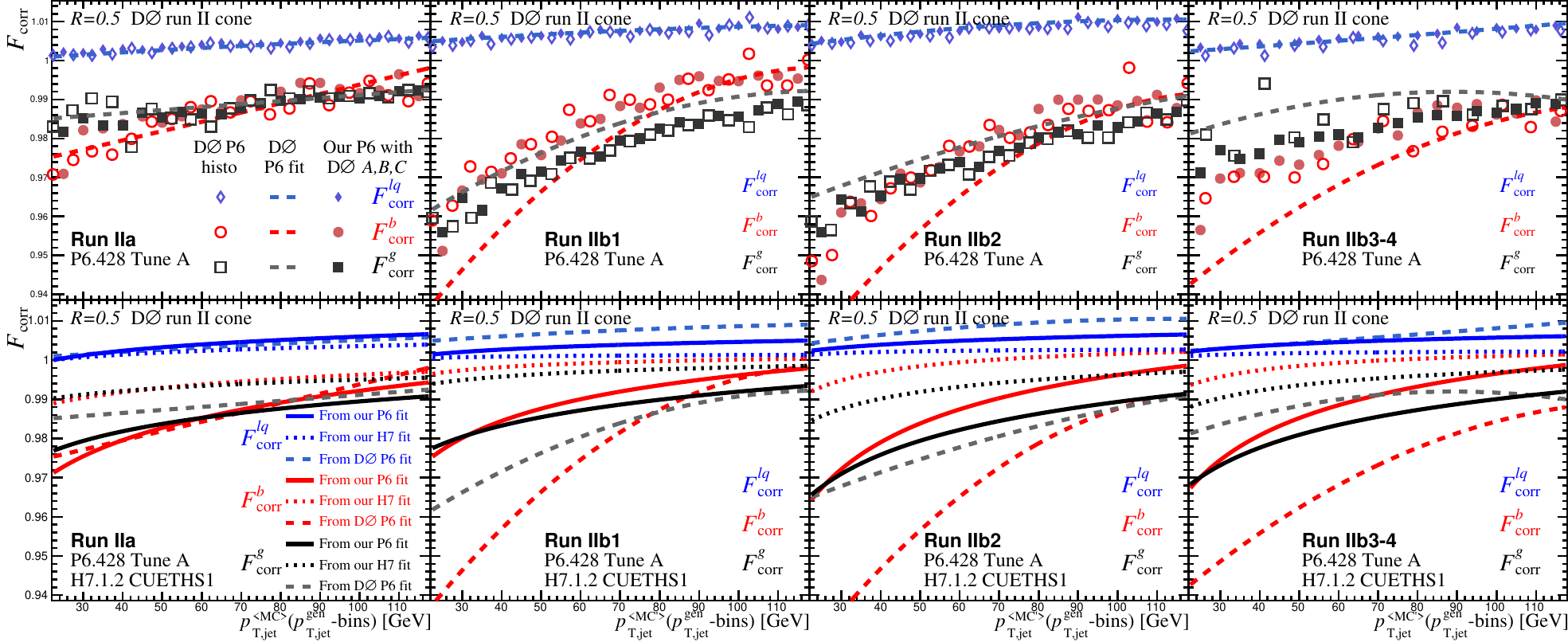}
\caption{
$F_{\textrm{corr}}$ values in the four D\O~RunII eras.
Row~1: D\O~histograms (open markers), D\O~fits (lines) and our P6 reproduction of the D\O~results (closed markers).
Row~2: our P6 (continuous line) and H7 (dotted line) fits vs. D\O~fits (dashed line).}
\end{figure}

\section{Observations and Interpretation}

Toni's analysis consists of two steps: refitting the 3 Single Particle Response (SPR) parameters and making parametrized $F_{\mathrm{Corr}}$ fits based on these.
The analysis was performed both on Pythia6 (P6) and Herwig7 (H7).
P6 was run with the exact same settings that D\O{} used.
H7 was used to probe the possible systematic errors, although no D\O{} tune was available.
We believe that Herwig or other secondary Monte Carlo (MC) event generators have not been utilized sufficiently in the
D\O{}'s Jet Energy Calibration and error analysis \cite{Abazov:2013hda}.

Toni's measurements are distilled into Figure~\ref{fig:toni}.
On Row 1, we observe D\O{}'s histogram-data and the fits based on these.
The closed markers display histograms based on Toni's SPR remakes.
It is observed that Toni has found the same histogram results as D\O{}.
In contrast, D\O{}'s histograms and fits agree only for Run~IIa, but not for the Run~IIb eras.
The histograms also show better time-stability than the fits.

On Row 2 of Fig.~\ref{fig:toni}, the histograms are not shown, but the same D\O{} fits are displayed.
In addition, Toni's fits are given.
His P6 fits agree with the histograms and are more stable in time than the D\O{} ones.
Both the fits and histograms for H7 show a great deviation from the P6 results.
The H7 histograms are not displayed in the Figure, but these agree with the H7 fits.
This motivates a suspicion of the remarkably small $F_{\mathrm{Corr}}$ errors stated by D\O{},
even if a D\O{} tune for H7 was not available.

The $m_t$ shifting method presented in Ref.~\cite{siikonen:2020mass} utilizes Toni's P6 and H7 fits.
The method takes D\O{}'s original fits as a reference and consequently probes the changes in the measured $m_t$ values introduced by Toni's fits.
The results are considered run-by-run (Run~IIa and Run~IIb\{1,2,34\}) and separating the electron and muon channels.
The $m_t$ analysis is performed purely with P6, utilizing D\O{}'s tuning.
We do not expect similar generator-dependence in the $m_t$ measurement, as in the $F_{\mathrm{Corr}}$ measurement.

\begin{figure} \label{fig:hannu}
\centering
\includegraphics[width=0.7\linewidth]{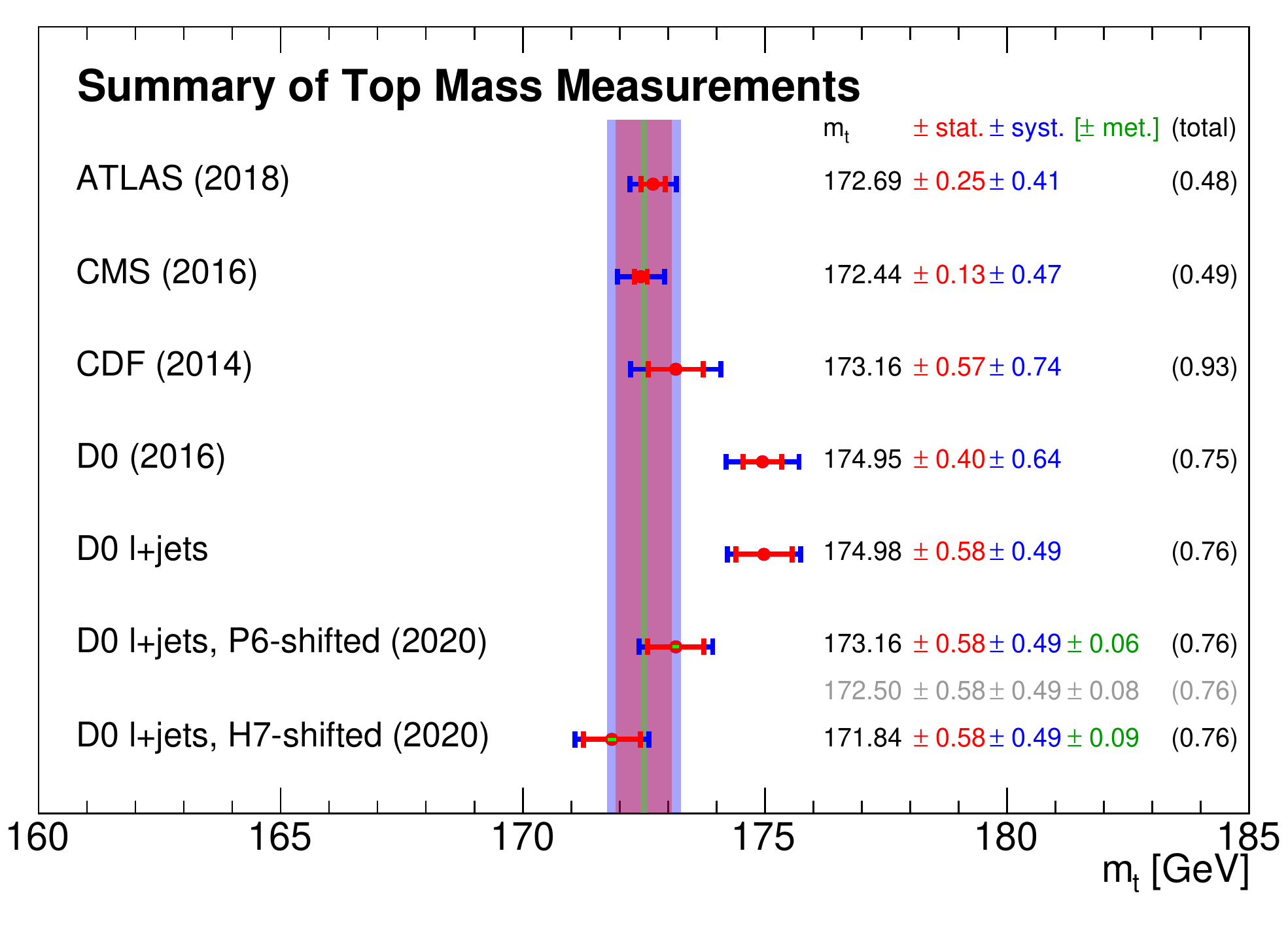}
\caption{The most prominent top mass measurements and the D\O~ measurements shifted using Toni's $F_{\textrm{corr}}$ values.
Lepton+jets channel dominates the D\O~result.}
\end{figure}

The final $m_t$ shifts using the P6 and H7 $F_{\mathrm{Corr}}$ values are given in Fig.~\ref{fig:hannu}.
It is noteworthy that the P6-shifted $m_t$ result is quite close to that of the CDF collaboration.
The generator choices of D\O{} and CDF are similar, and hence also the related systematic errors should agree.
The ATLAS and CMS results agree in a similar fashion as the CDF and the P6-shifted D\O{} results.
Interestingly, a simple linear combination of the P6-shifted and H7-shifted D\O{} results produces a value close to those of the CMS and ATLAS collaborations.
The LHC collaborations utilize more heavily Herwig in their jet calibrations than the Tevatron collaborations.
This could indicate that the choice of generators in the jet calibration process sets the rough baseline for results obtained in a precision measurement of the top mass.

\section{Discussion}

Toni has presented his results actively during 2018-2019, including a LHC Top WG meeting.
After finishing his Master's thesis, he left our group in mid-2019 to pursue a PhD in Germany.
The final $m_t$-shift results were published as an arXiv-preprint in early 2020, to be later included in the main author's PhD thesis.
In the spring of 2020, the results were presented and handled in an LHC Top WG meeting.
In this meeting, some of the D\O{} contacts got active, and were unknowing of Toni's previous contact efforts.
The situation was extraordinary, as Toni had been proactively in contact with the D\O{} authors.
The agreement of the meeting was that the mass shifting method itself seems trustworthy.
However, the D\O{} contacts disagreed on Toni's results.
Toni provided further information about his studies, but the contact from D\O{} dried up.

The $m_t$ shifting results were submitted as a poster to the TOP2020 conference.
We did not hear again from D\O{} sooner than one day before the poster presentation in a short letter \cite{ref:D0state}.
Here, it is addressed that \emph{There were no interactions between the authors of Refs. [1, 4] and the primary D0 authors concerning the details in the internal notes}.
This is not how we observed the process, where Toni has been very active into the direction of D\O{}.
Thanks to his contact with the D\O{} authors, he received all the material and calibrations reasonably available, making the former statement questionable.
Furthermore, some of the important findings were completely disregarded in the letter, such as the H7-based systematics treatment.

The letter aims for closing the discussion without needing to check the $F_{\mathrm{Corr}}$ calibration.
The disagreement between D\O{} fits and histograms is spoken off by mentioning that \emph{the plots in the internal notes could have been misleading}.
It is stated that the D\O{} review process is very thorough, and hence they do not expect any errors.
This is an incomplete argument for defending the calibration.
The final paper releases are less detailed than the analysis notes.
The details ending up in the published paper are generally reviewed with the help of the more detailed analysis notes.
If there are internal disagreements within the notes, how can the calibrations of the final paper be trusted?
To complete their argument, D\O{} should show that the final fits were indeed corrected from the ones presented in the note.
Without a visual or numerical confirmation, the statements of Ref. \cite{ref:D0state} remain unconvincing.

Despite the tight schedule, the TOP2020 organizers allowed adding a mention of the D\O{} letter into the poster.
Regrettably, a thorough reciprocal discussion of the D\O{} statements did not enter the poster session.
The audience was generally curious, and the most critical voices of the D\O{} letter were absent.

In short, these $m_t$ studies aimed for a D\O{} re-review of a specific part of their $F_{\mathrm{Corr}}$ calibration.
With the release of D\O{}'s internal notes, our collaboration with D\O{} started auspiciously.
Utilizing the available material, we have completed a detailed study of the D\O{} analysis chain.
However, D\O{} has responded to this only with a short letter including controversial statements.
This is not consistent with good scientific practice.
Going into more detail is infeasible without D\O{} taking the lead.

We wish that D\O{} would pick up the issue and re-review the $F_{\mathrm{Corr}}$ calibration.
We have pinpointed the location of the possible issues, so this should not be labor-intensive.
With the amount of work and evidence that we have provided,
failing to do this serves against the self-corrective ideals of science.


\begin{thebibliography}{99}


\bibitem{ref:d0comb}
  V.~M.~Abazov {\it et al.} [D0 Collaboration],
  Phys.\ Rev.\ D {\bf 95} (2017) no.11,  112004

\bibitem{pdg} K.~A.~Olive {\em et al.} (Particle Data Group), ``The Review of Particle Physics'', Chin.~Phys.~C, 38, 090001 (2014) and 2015 update.

\bibitem{ref:d02015}
  V.~M.~Abazov {\it et al.} [D0 Collaboration],
  Phys.\ Rev.\ D {\bf 91} (2015) no.11,  112003

\bibitem{ref:tonimaster}
T.~Mäkelä,
Aalto University Document Server (2019)
\url{https://aaltodoc.aalto.fi/handle/123456789/39024}

\bibitem{siikonen:2020mass}
H.~Siikonen,
2020-02-14,
\url{https://arxiv.org/abs/2002.06073}

\bibitem{Abazov:2013hda}
V.~M.~Abazov \textit{et al.} [D0],
Nucl. Instrum. Meth. A \textbf{763} (2014), 442-475

\bibitem{ref:D0state}
D\O{} collaboration,
\url{https://www-d0.fnal.gov/Run2Physics/WWW/results/final/TOP/T14E/D0_top-JES_statement.pdf}

\end{thebibliography}
\end{document}